\documentstyle[twoside]{article}
\catcode`\@=11
\long\def\@makefntext#1{
\protect\noindent \hbox to 3.2pt {\hskip-.9pt
$^{{\eightrm\@thefnmark}}$\hfil}#1\hfill}

\def\@makefnmark{\hbox to 0pt{$^{\@thefnmark}$\hss}}
\def\ps@myheadings{\let\@mkboth\@gobbletwo
\def\@oddhead{\hbox{}
\rightmark\hfil\eightrm\thepage}
\def\@oddfoot{}\def\@evenhead{\eightrm\thepage\hfil
\leftmark\hbox{}}\def\@evenfoot{}
\def\sectionmark##1{}\def\subsectionmark##1{}}
\oddsidemargin=\evensidemargin
\newcounter{sectionc}\newcounter{subsectionc}\newcounter{subsubsectionc}
\renewcommand{\section}[1] {\vspace{12pt}\addtocounter{sectionc}{1}
\setcounter{subsectionc}{0}\setcounter{subsubsectionc}{0}\noindent
    {\tenbf\thesectionc. #1}\par\vspace{5pt}}
\renewcommand{\subsection}[1] {\vspace{12pt}\addtocounter{subsectionc}{1}
    \setcounter{subsubsectionc}{0}\noindent
    {\bf\thesectionc.\thesubsectionc. {\kern1pt \bfit #1}}\par\vspace{5pt}}
\renewcommand{\subsubsection}[1] {\vspace{12pt}\addtocounter{subsubsectionc}{1}
    \noindent{\tenrm\thesectionc.\thesubsectionc.\thesubsubsectionc.
    {\kern1pt \tenit #1}}\par\vspace{5pt}}
\newcommand{\nonumsection}[1] {\vspace{12pt}\noindent{\tenbf #1}
    \par\vspace{5pt}}

\newcommand{\textlineskip}{\baselineskip=13pt}

\def\eightcirc{
\begin{picture}(0,0)
\put(4.4,1.8){\circle{6.5}}
\end{picture}}
\def\eightcopyright{\eightcirc\kern2.7pt\hbox{\eightrm c}}

\def\abstracts#1#2#3{{
    \centering{\begin{minipage}{4.5in}\footnotesize\baselineskip=10pt
    \parindent=0pt #1\par
    \parindent=15pt #2\par
    \parindent=15pt #3
    \end{minipage}}\par}}

\newcommand{\bibit}{\nineit}
\newcommand{\bibbf}{\ninebf}
\renewenvironment{thebibliography}[1]
    {\frenchspacing
     \ninerm\baselineskip=11pt
     \begin{list}{\arabic{enumi}.}
        {\usecounter{enumi}\setlength{\parsep}{0pt}
    \setlength{\leftmargin 17pt}{\rightmargin 0pt}   
         \setlength{\itemsep}{0pt} \settowidth
    {\labelwidth}{#1.}\sloppy}}{\end{list}}

\newcounter{itemlistc}
\newcounter{romanlistc}
\newcounter{alphlistc}
\newcounter{arabiclistc}

\def\@citex[#1]#2{\if@filesw\immediate\write\@auxout
    {\string\citation{#2}}\fi
\def\@citea{}\@cite{\@for\@citeb:=#2\do
    {\@citea\def\@citea{,}\@ifundefined
    {b@\@citeb}{{\bf ?}\@warning
    {Citation `\@citeb' on page \thepage \space undefined}}
    {\csname b@\@citeb\endcsname}}}{#1}}

\newif\if@cghi
\def\cite{\@cghitrue\@ifnextchar [{\@tempswatrue
    \@citex}{\@tempswafalse\@citex[]}}
\def\citelow{\@cghifalse\@ifnextchar [{\@tempswatrue
    \@citex}{\@tempswafalse\@citex[]}}
\def\@cite#1#2{{$\null^{#1}$\if@tempswa\typeout
    {IJCGA warning: optional citation argument
    ignored: `#2'} \fi}}

\def\pmb#1{\setbox0=\hbox{#1}
    \kern-.025em\copy0\kern-\wd0
    \kern.05em\copy0\kern-\wd0
    \kern-.025em\raise.0433em\box0}

\def\fnt#1#2{\footnotetext{\kern-.3em
    {$^{\mbox{\scriptsize #1}}$}{#2}}}

\def\fpage#1{\begingroup
\voffset=.3in
\thispagestyle{empty}\begin{table}[b]\centerline{\footnotesize #1}
    \end{table}\endgroup}

\font\tenrm=cmr10
\font\tenit=cmti10
\font\tenbf=cmbx10
\font\bfit=cmbxti10 at 10pt
\font\ninerm=cmr9
\font\nineit=cmti9
\font\ninebf=cmbx9
\font\eightrm=cmr8

\def\qed{\hbox{${\vcenter{\vbox{
   \hrule height 0.4pt\hbox{\vrule width 0.4pt height 6pt
   \kern5pt\vrule width 0.4pt}\hrule height 0.4pt}}}$}}

\def\be{\begin{equation}}
\def\ee{\end{equation}}

\begin{document}
\setlength{\textheight}{7.7truein}  


\normalsize\textlineskip
\thispagestyle{empty}
\setcounter{page}{1}

\vspace*{0.88truein}

\fpage{1}
\centerline{\bf CHIRAL STRING IN A CURVED SPACE:}
\baselineskip=13pt
\centerline{\bf GRAVITATIONAL SELF-ACTION}

\vspace*{0.37truein} \centerline{\footnotesize Yu.V. GRATS
\footnote{E-mail address: grats@string.phys.msu.su},  A.A.
ROSSIKHIN and A.O. SBOICHAKOV}
\baselineskip=12pt
\centerline{\footnotesize\it Department of Theoretical Physics,
M.V. Lomonosov Moscow State University}
\baselineskip=10pt
\centerline{\footnotesize\it 119899, Moscow, Russia}

\vspace*{0.21truein}
\abstracts{
We analyze the effective action
describing the linearised gravitational self-action for a classical
superconducting string in a curved spacetime.  It is shown that the
divergent part of the effective action is equal to zero for the both
Nambu-Goto string and chiral superconducting string.
}{}{}

\textlineskip
\vspace*{12pt}
\noindent
Now it is well understood that topological defects in quantum field
theory may play an essential role in cosmology, and the vertex
line defects describable on a macroscopic scale as cosmic strings are
most likely to be actually generated at phase transitions in the early
Universe.$^1$ Strings are predicted by many of the commonly
considered field theoretical models, and their gravitational effects
become the question of increasing interest during last twenty years.
Indeed, the evolution of a cosmic string network may play an essential
role in the formation of the large scale structure of the Universe
observed at the present time. But possible cosmological
applications is not the only reason for this consideration. There are some
nontrivial features in the interaction of strings with gravity.
One of them is the absence of gravitational bremsstrachlung from
colliding straight gravitating segments$^2$, while test particle freely
moving near Nambu-Goto string emits gravitational waves.$^3$ Another
one is the absence of classical linearised self-interaction of a
Nambu-Goto string with a gravitational field in a four-dimensional
Minkowski spacetime.$^{4, 5}$ The last result is nontrivial.
We know that in a lot of applications the radius of curvature of a
string is much larger that its transverse diameter. So, the string can
be considered to be infinitely thin, and the dynamics of the vertex
line can be described with the use of the Nambu-Goto action or its
generalization, if the interaction with an electromagnetic field or
the fields of axion and dilaton is taken into account.  In these
cases underlying gauge theory is used only for the fixing the
parameters in a string action.  Interaction with matter
fields give rise to outgoing flaxes of corresponding quanta. At the
same time zero thickness of strings
must lead to the divergence of a self-force.  This problem is
well known in classical electrodynamics. In the case of point charge in
Minkowski space such a situation was considered by Dirac$^6$ who showed
that the divergence in the self-action can be removed by the
renormalization of the particle mass with the use of corresponding
classical counterterm.  Analogous problem has been studied in a lot of
papers for the string interacting with linearised gravitational field
on Minkowski background, with electromagnetic field, with dilaton and
axion, see [4, 5] and the Refs.  therein.  From our point of view, the
absence of classical linearised gravitational self-interaction is the
most interesting result of the previous consideration.  In the case
of straight Nambu-Goto string this conclusion is more or less
obvious because Newtonian potential of such a string is equal to zero.
But for the curved string, for the string in an external gravitational
field or for the string with some internal degrees of freedom, say a
currentcarrying one, the answer is not obvious.  So, we can say that
even at classical level interaction of cosmic strings with
gravitational field, and in particular gravitational self-interaction,
is not properly investigated.

Here we consider gravitational self-action of a chiral currentcarrying
string, i.e. a superconducting string with an isotropic current.
Chiral strings are of increasing interest because
this simple model may be considered as a first step towards the
understanding of a full theory including superconductivity in strings.
>From the other hand, vertex defects of this type arise
naturally in some kinds of supersymmetric theory, and
there are some interesting cosmological consequences of
their existence. In particular, current can stabilize cosmic string,
and chiral vortons are expected to be more stable than non-chiral
string loops.

We use the metric with the signature $(+---)$ and the system of units
$c=1$.

In order to describe the macroscopic
effects of the currents arising from the processes of
superconducting kind, Witten$^7$ proposed the use of simple
generalization of the Nambu-Goto model which is characterized by
the action
\be
\label{a1}
S_{str}=\int d^2\zeta\sqrt{-\gamma}\Biggl(-\mu +
\frac{1}{2}\gamma^{ab}
\partial_a\phi\partial_b\phi\Biggr)\ ,
\ee
where $\zeta^a\ a=0, 1$ are the internal coordinates on
the world sheet which is imbedded in the four-dimensional spacetime with
coordinates $x^{\mu}$ and metric $g_{\mu\nu}$, $\gamma$ is the
determinant of the induced metric
$\gamma_{ab}=g_{\mu\nu}x^{\mu}_{,a}x^{\nu}_{,b}$.

In this model there is an additional internal scalar field $\phi$
in terms of which one can express the worldsheet supported
conserved current
\be
\label{a2}
{\cal J}^a=\frac{1}{\sqrt{-\gamma}}e^{ab}\phi_{,b}
\quad {\cal J}^a{}_{;a}=0
\ee
and energy-momentum tensor
\be
\label{a3}
{\cal T}_{ab}=\mu\gamma_{ab}+\phi_{,a}\phi_{,b}-
\frac{1}{2}\gamma_{ab}\phi_{,c}\phi^{,c}\ .
\ee
In Eq.~(3) $e^{ab}/\sqrt{-\gamma}$ stands for the two-dimensional
Levi-Civita tensor, with $e^{01}=-e^{10}=-1$.

The surface current ${\cal J}^a$ and energy-momentum tensor ${\cal
T}_{ab}$ are the two-dimensional tensorial fields defined on
the worldsheet of a string, while corresponding four-dimensional
sources which determine electromagnetic and gravitational interaction of
the string read
$$
J^{\mu}(x)=\int d^2\zeta\sqrt{-\gamma}{\cal J}^a
\partial_a x^{\mu} \delta(x, x(\zeta))\ ,
$$
\be
\label{b4}
T^{\alpha\beta}(x)=\int d^2 \zeta\sqrt{-\gamma}{\cal
T}^{ab}(\zeta)\partial_{a}x^{\alpha}\partial_{b}x^{\beta}
\delta^4(x, x(\zeta))\ .
\ee
And, as in the Dirac's case, the ultraviolet divergence in a
self-force, if it is, is the consequence of a distributional nature
of the currents ~(4).

Below we will
consider the case of chiral string with a so-called null current.
This condition does not mean that the current itself must be equal to
zero, but that it is timelike
\be
\label{a5}
\gamma_{ab}{\cal J}^a {\cal J}^b =0=
\gamma^{ab}\phi_{,a}\phi_{,b}\ .
\ee
For the chiral string the last term in Eq.~(3) vanishes, but
the energy-momentum tensor still differs from the one describing string
of the Nambu-Goto type. So, one can suppose the existence of some
differences in the interaction of a Nambu-Goto and currentcarrying
strings with an external gravitational field.

To allow for the interaction with a linearised gravitational
perturbations, one must perform the replacement $g\mapsto g+h$ which
is equivalent to the augmentation of the string Lagrangian in Eg.~(1) by
the interaction term
\be
\label{a6}
{\cal L}_{str}[g]\mapsto {\cal L}_{str}[g] - \frac{1}{2}{\cal T}^{ab}
\partial_a x^{\mu}\partial_b x^{\nu}h_{\mu\nu}(x(\zeta))\ ,
\ee
As for the gravitational action, we must expand it in powers of
$h_{\mu\nu}$ up to the second-order terms.

Corresponding equation for the linear metric perturbations has the
wellknown form
\be
\label{a7}
\delta\frac{\delta S_{gr}[g]}{\delta
g^{\mu\nu}(x)}=-\frac{\sqrt{-g}}{2}T_{\mu\nu}(x)\ .
\ee
Second-order variational derivative of the gravitational action can be
found for example in Ref.[7].

Substituting the solution of the Eq.~(7) back into the total action, we
obtain
$$
S_{tot}[g, x, \phi]=S_{gr}[g]+S_{str}[g, x, \phi]+
$$
\be
\label{a8}
+\frac{G}{4}\int
d^2\zeta\sqrt{-\gamma}\int d^2\zeta'\sqrt{-\gamma'}
{\cal T}^{\mu\nu}(\zeta){\bar G}_{\mu\nu\mu'\nu'}^{gr}(\zeta,\zeta')
{\cal T}^{\mu'\nu'}(\zeta')\ ,
\ee
where for simplisity we use the notation
${\cal T}^{\mu\nu}=\partial_a x^{\mu}\partial_b x^{\nu}{\cal T}^{ab}$.

The second and the third terms in the Eq.(8) depend on the string
variables and describe the dynamics of an infinitely thin
cosmic string which interacts with it's linearised gravitational field.
When gravitational radiation from the string is not taken into
account $\bar G_{\mu\nu\mu'\nu'}^{gr}$ is the
real part of the Feynman propagator.

In the Lorentz gauge,
when $(h_{\mu\nu}-1/2g_{\mu\nu}h)^{;\mu}=0$, this equation
becomes
\be
\label{a9}
\left( \delta^{\alpha}_{\mu}\delta^{\beta}_{\nu}\Box +
H_{gr\ \mu\nu}^{\
\alpha\beta}\right)G^{gr}_{\alpha\beta\mu'\nu'}(x, x')
=-16\pi\left( g_{\mu\mu'}g_{\nu\nu'}-\frac{1}{2}
g_{\mu\nu}g_{\mu'\nu'}\right)\delta(x, x')\ ,
\ee
and $H_{gr\ \mu\nu}^{\ \alpha\beta}$ takes the form
\be
\label{a10}
H_{gr\ \mu\nu}^{\ \alpha\beta}=-2R^{\alpha\ \beta}_{\
\mu\nu}-2\delta^{\alpha}_{(\mu}R^{\beta}_{\nu)}-\frac{1}{2}g_{\mu\nu}
g^{\alpha\beta}R+
g_{\mu\nu}R^{\alpha\beta}+\delta^{\alpha}_{\mu}\delta^{\beta}_{\nu}R\ .
\ee

Proceeding along the same line as in Ref.~[9] one can obtain
the expression
$$
\bar G^{gr}_{\mu\nu\mu'\nu'}(x, x')=2\Delta^{1/2}(x,
x') \left(\epsilon_{\mu\mu'}\epsilon_{\nu\nu'}-
\frac{1}{2}g_{\mu\nu}g_{\mu'\nu'}\right)
\delta(\sigma)+
$$
\be
\label{b11}
+\theta(-\sigma)v_{\mu\nu\mu'\nu'}(x, x')\ .
\ee
In the last equation $\Delta(x,
x')=-(-g(x))^{-1/2}\det\left(\partial^2_{\mu\nu'} \sigma(x,
x')\right)(-g(x'))^{-1/2}$,  $\epsilon_{\mu\mu'}$ is the
bivector of parallel transport, $\sigma$ is the half of geodesic
distance between points $x$ and ${x'}$, and $v_{\mu\nu\mu'\nu'}(x, x')$
denotes some regular function, which must be calculated for any
particular case.

>From Eq.~(11) we can see that the part of the action we would expect to
be divergent is
$$
S_{div}=\frac{G}{2}\int d^2\zeta \sqrt{-\gamma(\zeta)}\int d^2 \zeta'
\sqrt{-\gamma(\zeta')} \Delta^{1/2}(\zeta,\zeta')
\delta\left(\sigma(\zeta,\zeta')\right)\times
$$
\be
\label{a12}
\times{\cal T}^{\mu\nu}(\zeta)
\left(\epsilon_{\mu\mu'}\epsilon_{\nu\nu'}-
\frac{1}{2}g_{\mu\nu}g_{\mu'\nu'}\right)
{\cal T}^{\mu'\nu'}(\zeta')\ .
\ee
In four spacetime dimensions the integral in Eq.(12) diverges
logarithmically as $\zeta\to\zeta'$, and it must be renormalized.
This integral can be estimated with the use of the normal Riemann
coordinates on the world sheet with the origin at the point $\zeta$ or
using any other regularization scheme, see for example Ref.~[5],
\be
\label{a13}
S_{eff}=2G\log \frac{\triangle}{\delta}\int d^2
\zeta\sqrt{-\gamma}{\cal T}^{\mu\nu}\left(
g_{\mu\gamma}g_{\nu\delta}-
\frac{1}{2}g_{\mu\nu}g_{\gamma\delta}\right) {\cal
T}^{\gamma\delta}+S_{fin}\ ,
\ee
where $\delta$ is a short-range cutoff length, which is identified
with the string radius, while $\Delta$ must be introduced because of
the logarithmic dependence of $\delta$ in Eq.~(13). This infrared
regularization length corresponds to the distance at which conical
geometry of string is disturbed by the background geometry.

Now we have to substitute the explicit expression for ${\cal
T}^{\mu\nu}$ into Eq.~(13). We then obtain
\be
\label{a14}
S_{div}=G\log\frac{\Delta}{\delta}\int
d^2\zeta\sqrt{-\gamma}{\cal J}^4\ .
\ee

We calculated the divergent part of the effective action for a
currentcarrying string interacting with gravitational field.
Eq.~(14) shows that linearised gravitational self-interaction in the
presence of an external gravitational field is equal to zero for the
both Nambu-Goto and superconducting chiral string.
This result generalizes the one obtained in Refs.~[4, 5].

The finite part of the action determines the effect of the topological
self-action which is equal to zero in the Minkowski space. There is no
general prescription for its calculation.  Sometimes the finite part of
the self-force can be calculated explicitly$^{10, 11}$ or with the use
of perturbative technique.$^{12}$

If ${\cal J}^2\neq 0$ the situation becomes some more interesting. In
this case the logarithmic divergent term ~(14) does not renormalize the
bare action ~(1) because there is no term $\sim {\cal J}^4$ in Eq.~(1).
It seems that this seems that the full noncontradictional theory of
superconducting string must be a nonlinear theory.

\nonumsection{Acknowledgments}
\noindent
This work was supported by the Russian Foundation for
Basic Research, grant 99-02-16132.

\nonumsection{References}

\end{document}